\documentstyle[aps,epsfig,osa]{revtex}
\begin{document}
\title{Mean-Field-Theory for Polymers in Mixed Solvents
Thermodynamic and Structural Properties}
\author{Amina Negadi$^{1,2}$, Anne Sans-Pennincks$^{3}$, M. Benmouna$^{1,2}$, 
and Thomas A. Vilgis*$^{1,4}$}
\address{$^{1}$Max-Planck Institut f\"ur Polymerforschung,
 Ackermannweg 10, D-55122  Mainz, Germany}
\address{$^{2}$ University Aboubakr Belkaid of Tlemcen, 
Institut of 
chemistry, BP 119, Tlemcen 13000, Algeria}
\address{$^{3}$ Centre d'etudes de Bruyeres le Chatel, 
B.P. 12; 
F-91680 Bruyeres les Chatel, France}
\address{$^{4}$ Laboratoire Europ\'een Associ\'e, ICS, 6, rue Boussingault,
  F-67083 Strasbourg, France}
\date{\today}
\maketitle
\begin{abstract}
Theoretical aspects of polymers in mixed solvents are considered
using the Edwards Hamiltonian formalism. Thermodynamic and structural
properties are investigated and some predictions are made when the mixed solvent approaches criticality. Both the single and the many chain problems
are examined. When the pure mixed solvent is near criticality,
addition of a small amount of polymers shifts the criticality towards
either enhanced compatibility or induced phase separation 
depending upon the value of the parameter describing the interaction
asymmetry of the solvents with respect to the polymer. The
polymer-solvent effective interaction parameter increases strongly
when the solvent mixture approaches criticality. Accordingly, the apparent
excluded volume parameter decreases and may vanish or even become negative. 
Consequently, the polymer undergoes a phase transition from a
swollen state to an unperturbed state or even take a collapsed configuration. 
The
effective potential acting on a test chain in strong solutions is calculated
and the concept of Edwards screening discussed. Structural properties
of ternary mixtures of polymers in mixed solvents are investigated within the
Edwards Hamiltonian model. It is shown that the effective potential on a test chain in strong
solutions could be written as an infinite series expansion of terms describing
interactions via one chain, two chains etc. This summation can be
performed following a similar scheme as in the Ornstein-Zernike series
expansion.
\end{abstract}

\section{Introduction}
Polymers in mixed solvents show different properties than in individual 
solvents raising interesting questions from a fundamental point of view
\cite{Schultz:55.1,Dondos:70.1,Brochard:80.1,Magda:88.1,Vilgms:93.1,Vilgms:98.1}.
Specific features of these systems originate from incompatibility of the two solvents and asymmetry of their interactions with the
polymer. If the system exhibits a
preferential adsorption of one solvent with the polymer, a
difference of solvent composition inside and outside the polymer domain is 
observed
\cite{prefads:77.1,prefads:78.1,prefads:79.1,prefads:81.1,prefads:88.1}.
Another peculiar behavior of dilute solutions of
polymers in binary mixtures of bad solvents is 
cosolvency \cite{Cosolv:81.1,Cosolv:85.1,Cosolv:88.1,Cosolv:96.1}
This behavior is quite unexpected since two non solvents
for the polymer turn out to present together the properties of a good solvent
which is unusual from the  point of view of
excluded volume effects. If no specific interactions exist between particles
in the mixture such as hydrogen bonds or long range
electrostatic interaction, one would expect that a mixed solvent
exhibits a smooth transition from one solvent quality to the other as
the solvent composition changes from 0 to 1 and the solvent power
should be an averaged value of its individual components. This is not the case in
cosolvent systems where sharp changes characteristic of first order
transitions take place. These transitions can be monitored through
variations of radii of gyration, second virial
coefficients, solution viscosities ... with the solvent composition.

In recent years, with the development of new techniques of polymer characterization
using molecular labeling, dyes and neutron scattering techniques,
...widespread  research of polymers in mixed solvents has been witnessed
in an attempt to understand better their fundamental
properties.
These mixtures were
studied by many authors using different methods such as light and
neutron scattering. 
Mixtures of ordinary and deuterated solvents such as light and heavy
water, ordinary and deuterated organic solvents
(benzene, toluene ...) are used quite often in neutron scattering
measurements to achieve maximum contrast taking advantage of the large
difference between the coherent scattering lengths of hydrogen and deuterium
atoms \cite{Higgins:94.1}.   

In polyelectrolyte solutions, a low molecular weight salt is added to increase
the ionic strength which weakens the electrostatic forces and
lowers  their
range. Salt introduces two additional low molecular weight components
(co-ions and counter-ions) which could present the properties of mixed
solvents \cite{Norwood:96.1}.

Recently, mixed solvents were used to study the thermodynamics of polymer
brushes \cite{Auroy:92.1,Lyatskaya:97.1,Ruhe:98.1}. Auroy and
Auvray\cite{Auroy:92.1} observed the collapse-stretching transition of
grafted polydimethylsiloxane chains by small angle neutron scattering. Changing temperature
and composition, they were able  to control the height of the
brush. The special boundary
conditions in the terminal attachments result into polymer conformation
and dynamics which are different from those of the
bulk systems. The
degree of swelling of polymer brushes is largely determined by the
solvent quality. Rothers et al.\cite{Ruhe:98.1} investigated the swelling of
polystyrene (PS) brushes in toluene-methanol mixtures. By changing the solvent composition,
they were able to see the collapse-stretching transition in the brush.

Problems involving polymers in mixed solvents are not only of fundamental
interest but also they raise interesting concrete questions and offer a
large spectrum of practical applications. The increase of amount  of water into polymer
domains in nylon technology can be monitored directly  by NMR imaging
techniques and give direct profiles of solvent concentration and ingress dynamics
\cite{Ingress:92.1,Ingress:97.1}.
Research efforts are engaged by polymer industries and rayon manufacture to
develop new clean methods to meet the requirements of environmental
regulations \cite{Morgenstern:96.1}. In these efforts applications of mixed solvent systems are often
considered. Polymer synthesis in mixed solvents is another application which is
found to allow for a good control of the polymerization process
\cite{synthese:96.1,synthese:97.1}.

In this paper, we are interested in the properties of polymers in mixed
solvents from the fundamental point of view and present a theoretical
investigation based upon the Edwards Hamiltonian formalism\cite {Doi:86.1}. 
In the next section, we review the case of a single chain in mixed
solvents. The effective potential acting on the chain is calculated together with
the scattering functions. Phase behavior and chain conformation are obtained
from the generalized partition function in the Edwards Hamiltonian formalism. In section 3, we consider the case of strong
solutions and discuss changes in the phase behavior and the scattering
properties due to the polymer concentration of the medium. Section 4 gives concluding remarks.

\section{The single chain in mixed solvents}

\subsection{The Edwards Hamiltonian formalism}
In this section, we consider infinite dilute
solutions of polymers in mixed solvents and use the Edwards Hamiltonian
formalism to determine the  effective potential
exerted on a chain under various conditions of temperature and
solvent composition. For simplicity, we assume that all the chains have the
same degree of polymerization $N$ and the same length $l$ to avoid
complications due to polydispersity. For this idealized system, the Edwards Hamiltonian is
\begin{eqnarray}
  \beta {\cal H} & = & \frac{3}{2 l^{2}}\int_0^N \left(\frac {\partial{\bf R}(s)}{\partial
    s}\right)^{2} 
ds +\frac{1}{2} \int_0^N\int_0^N V_{\rm {PP}}[({\bf R}(s)-{\bf R}(s')]ds ds' \nonumber \\ & + &
\sum_{i}\int_0^N V_{\rm {AP}}[{\bf R}(s)-{\bf r}_i^A]ds 
+\sum_{i}\int_0^N V_{\rm {BP}}[{\bf R}(s)-{\bf r}_i^B]ds \nonumber \\ & + &
\frac{1}{2}\sum_{i,j}^N V_{\rm {AA}}
[{\bf r}_i^A-{\bf r}_j^A]+\frac{1}{2}\sum_{i,j} V_{\rm {BB}}[{\bf r}_i^B -
{\bf r}_j^B]+\sum_{i,j}V_{\rm {AB}}
[{\bf r}_i^A-{\bf r}_j^B]
\end{eqnarray}
where $\beta = 1/k_{\rm B}T$, $k_B$ is the 
Boltzmann constant and T the absolute
temperature, ${\bf R}(s)$ is the chain variable and $s$ the curvilinear
coordinates along the chain ( contour variable). The first term in the right
hand 
side (RHS) of equation (1) is the
Wiener measure for random walks (assuming that unperturbated chains are Gaussian) and represents the chain entropy
due to its elasticity. The other terms in the RHS of equation 1 result from 
two-body interactions between various particles in the medium either of the same or different species. In this expression, we
have used the assumption that the interaction potentials are
short ranged and can be approximated with Dirac delta functions

\begin{equation}
V_{ij}({\bf {r}_i} - {\bf {r}_j})= V_{ij} \delta ({{\bf r}_i} -  {{\bf r}_j})
\end{equation}
where $ V_{ij}$ is the strength of the interaction potential between
particles $i$ and $j$. Using the standard procedures of functional theory and
Edwards Hamiltonian formalism, we derive the thermodynamic and the structural
properties of the mixture under consideration from the generalized partition
function. Starting from the Hamiltonian of $equation$ $ 1$, one obtains the
partition function as a functional integral over all monomers and
solvent coordinates

\begin{equation}
{\cal Z} = \int {{\bf \cal
  D}}{R(s)}\int\prod_{i} d{{\bf r}_i^A}\prod_{i}
d{{\bf r}_i^B}\exp{-\beta{\cal H}[({\bf R}(s),{\bf r}_i^A,{\bf r}_i^B]}
\end{equation}
Where the superscripts $A$ and $B$ refer to the solvents. It is quite difficult to handle the partition function in the
particle position variable. Rather, it is more convenient to
transform into collective variables such as particle densities
\begin{eqnarray}
\rho^A_{\bf q}= \sum_i\exp (- \rm i{\bf q \bf r}_i^A) \nonumber\\
\rho^B_{\bf q}= \sum_i\exp (- \rm i{\bf q \bf r}_i^B)
\end{eqnarray}
Considering the single chain in solution and integrating over solvent 
coordinates, one obtains the Edwards Hamiltonian for the chain subject to the
effective potential $V_{{0}}({\bf q})$ 
\begin{equation}
\beta {\cal H} = \frac{3}{2l^2}\int_0^N \left( \frac{\partial {\bf R}(s)}{\partial s}\right)^{2}ds
+ \sum_{{\bf q}} \int_0^N ds \int_0^N ds' V_{0}({\bf q}) \exp [- \rm i{\bf q} ({\bf R}(s)-{\bf R}(s'))]
\end{equation}
The explicit form of $V_{0}({\bf q})$ is obtained following
the procedure described below.

\subsection{The effective potential of a single chain $ V_0({\bf q})$}
The partition function in equation (3) can be written differently by making
use of the collective coordinates $\rho_{\bf q}^{\rm A}$ and $\rho_{\bf
  q}^{\rm B}$ 
\begin{eqnarray}
{\cal Z} = \int {\cal D} {\bf R}(s)\int\prod_{\bf q} 
\delta \rho_{\bf q}^{\rm A}
\prod_{\bf q}
\delta \rho_{\bf q}^{\rm B} 
\exp
[
- \frac {3}{2l^{2}}\int_{0}^{N} 
\left( \frac {\partial{\bf R}(s)} {\partial s}
\right)^{2} ds - \frac{1}{2} V_{\rm PP}\int_{0}^{N}\int_{0}^{N} ds
ds' \delta[{\bf R}(s)-{\bf R}(s`)]\nonumber\\  
 - V_{\rm AP} \sum_{\bf q} \int_{0}^{N}\exp (- \rm i{\bf q R}(s)) 
\rho_{\bf q}^{\rm A}
 - V_{\rm BP} \sum_{\bf q} \int_{0}^{N}\exp (- \rm i{\bf q R}(s)) 
\rho_{\bf q}^{\rm B} \nonumber\\  -
  \frac{1}{2} V_{\rm AA} \sum_{\bf q} |\rho_{\bf q}^{\rm A}|^{2} - 
\frac{1}{2}
  V_{\rm BB} \sum_{\bf q} |\rho_{\bf q}^{\rm B}|^{2} - V_{\rm AB} 
\sum_{\bf q} \rho_{\bf
    q}^{\rm A}\rho_{\bf -q}^{\rm B} 
]
\end{eqnarray}
where the letter $P$ refers to the polymer. The 
particle conservation can be
expressed by using the following equation
\begin{equation}
\int \prod_{\bf q}\delta \rho_{\bf q} \delta \left[\rho_{\bf q} - \sum_{\bf i} \exp( - \rm i{\bf q r}_{i})\right] = 1
\end{equation}
and introducing field variables via the auxiliary Edwards random fields
$\Phi$ 
\begin{equation}
\delta [\rho_{\bf q} - \sum_{i} \exp( - \rm i{\bf q r}_{i})] = \int \delta \Phi_{\bf
  q} \exp{[ - \rm i\Phi_{\bf q} \rho_{\bf -q}]}
\exp{[ - \rm i\Phi_{\bf q} \sum_{i}  - \rm i{\bf q r}_{i}]}
\end{equation}
With straightforward mathematical manipulations, one arrives at
\begin{eqnarray}
{\cal Z} = \int {{\bf \cal D}} {\bf R} (s)\int\prod \delta \rho_{\bf q}^{\rm A}
\prod \delta \rho_{\bf q}^{\rm B} \exp{[ - \frac {3}{2l^{2}} \int_{0}^{N} \left(
    \frac {\partial{\bf R}(s)} {\partial s}\right) ^{2} ds - \frac{1}{2}
  V_{\rm PP}\int_{0}^{N}\int_{0}^{N} ds}
 ds' \delta [{\bf R}(s)-{\bf R}(s')] \nonumber\\ - V_{\rm AP} \sum_{\bf q}
 \int_{0}^{N} ds \exp( -\rm i {\bf q R}(s)) \rho_{\bf q}^{\rm A}
  - V_{\rm BP} \sum_{\bf q}
  \int_{0}^{N} ds \exp (-\rm i {\bf q R}(s)) \rho_{\bf q}^{\rm B} \nonumber\\  -
  \frac{1}{2} \sum_{\bf q} (V_{\rm AA} +
  \frac {1}{S_{\rm A}}) |\rho_{\bf q} ^ {\rm A}| ^{2} - 
\frac{1} {2} \sum_{\bf q} (V_{\rm BB} + 
 \frac{1}{S_{\rm B}}) |\rho_{\bf q}^{\rm B}|^{2} - V_{\rm AB} \sum_{\bf q}
    \rho_{\bf q}^{\rm A}\rho_{\bf -q}^{\rm B}] 
\end{eqnarray}

For an incompressible mixture, the total mean density fluctuation is zero and
one has
\begin{equation}
\rho_{\bf q}^{\rm A}+ \rho_{\bf q}^{\rm B} + \int_{0}^{N} ds 
\exp ({-\rm i{\bf q R}(s))} = 0
\end{equation}
Substituting this result into equation (9) and integrating over solvent
collective coordinates, one obtains the partition function in terms of an
effective Hamiltonian ${\cal H}$ given by
\begin{equation}
\beta {\cal H} = \frac{3}{2l^{2}}\int_0^N \left( \frac {\partial {\bf
      R}(s)}{\partial s}\right)^{2} ds
    +  \int_0^N \int_0^N V_{0}({\bf q})
\exp[ - {\rm i}{\bf q} ({\bf R}(s)-{\bf R}(s'))] ds ds'
\end{equation}
where the effective potential $ V_{0}({\bf q})$ is function of the potentials
$V_{ij}$ and the partial structure factors $S_{\rm A}({\bf q})$ and $S_{\rm B}({\bf q})$
\begin{eqnarray} 
V_{0}({\bf q})= V_{\rm PP} + V_{\rm BB} + \frac {1}{S_{\rm B}({\bf q})} 
- 2 V_{\rm BP}-
\frac {\left (V_{\rm BB} - V_{\rm AB} + V_{\rm AP} - V_{\rm BP} +
    \frac{1}{S_{\rm B}({\bf q}) }\right)^{2}} {V_{\rm AA} +
V_{\rm BB} - 2 V_{\rm AB} + \frac {1}{S_{\rm A}({\bf q}) } + 
\frac {1}{S_{\rm B}({\bf q})} } 
\end{eqnarray}
The subscripts of potentials run over polymer $P$ and solvents $A$ and
$B$. To
make contact with measurable quantities, it is convenient to express
$V_{0}({\bf q})$ in terms 
of the Flory-Huggins interaction parameter $\chi _{ij}$ \cite{Flory:53.1} 
\begin{equation}
\chi _{ij} = V_{ij} - \frac{V_{ii} + V_{jj}}{2} 
\end{equation}
Combining equations (12) and (13) yields
\begin{equation}
V_{0}({\bf q}) = \frac{S_{\rm A}^{-1}({\bf q}) S_{\rm B}^{-1}({\bf q}) - 2
  \chi_{\rm AP}
  S_{\rm B}^{-1}({\bf q})  - 2 \chi_{\rm BP} S_{\rm A}^{-1}({\bf q})
  + D}{S_{\rm A}^{-1}({\bf q}) + S_{\rm B}^{-1}({\bf q}) - 2 \chi_{\rm AB}}
\end{equation}
where $D$ is a function of the corresponding Flory-Huggins interaction \cite{Schultz:55.1} parameters only
\begin{equation}
D = 2 \chi_{\rm AB}\chi_{\rm AP} + 2 \chi_{\rm AB}\chi_{\rm BP} + 2 \chi_{\rm
  AP}\chi_{\rm BP}-
\chi_{\rm AP}^{2} - \chi_{\rm BP}^{2} - \chi_{\rm AB}^{2}
\end{equation}
Equation (14) could be written in a different form which is appealing from
the point of view of the effective potential acting on a polymer chain in mixed
solvents
\begin{equation}  
V_{0}({\bf q}) = \frac{1}{S_{\rm B}({\bf q})}- 2 \chi_{\rm BP}^{'} =
\frac{1}{S_{\rm B}({\bf q})} - 2 \chi_{\rm BP}-
\frac{\left ({S_{\rm B}^{-1}({\bf q}) - \chi_{\rm AB} - \chi_{\rm BP} +
      \chi_{\rm AP}}\right
  )^{2}}{S_{\rm A}^{-1}({\bf q}) +
  S_{\rm B}^{-1}({\bf q}) - 2 \chi_{\rm AB}}
\end{equation}

Similar results were obtained by others and in particular by Schultz and
Flory \cite{Schultz:55.1} and by Benoit and Strazielle \cite{Benoit:95.1} using
different methods of the mean field theory. The 
Edwards Hamiltonian method gives
another route for deriving the 
effective potential on a chain together with the scattering properties. It has
the merit of showing the response of a polymer chain undergoing interaction from the mixed solvent under various
conditions of temperature, composition, etc. 
\subsection{The scattering functions}

In the following subsection  the crudest approximation that  can be
imagined. Although we consider only one chain in the mixed solution we use a
mean field description of the RPA form. We are aware that this approximation
is very bad indeed, nevertheless it is used here to see the principal influence
of the binary solvent on the interaction potentials. Has we used Gaussian
chains this approximation would be fine. This case belongs to $V_{\rm PP}=0$.
Another possibility of slight improvement is to treat the polymer plus
excluded volume interaction as effectively "bare" system and
use an appropriate Pad\'e approximation including the selfavoiding walk exponent
$\nu \approx 3/5$.

Starting from equation 11, after some straightforward manipulations, one could write
the Edwards Hamiltonian in terms of the scattering functions for
incompressible mixtures of polymers and mixed solvents. The following
conventional notation is adopted. For interaction potentials, we keep the subscript $P$
for all the polymer chains but for the single chain structure factor we use the subscript
$1$ meaning a single test chain. With this convention, the Edwards Hamiltonian becomes 
\begin{eqnarray}
\beta {\cal H} = \sum_{\bf q}  \left [ S_{1}^{-1}({\bf q}) + V_{\rm PP} + V_{\rm BB}
+ S_{\rm B}^{- 1}({\bf q}) - 2 V_{\rm BP} \right]
|\rho_{\bf q}^{1}|^{2} \nonumber\\+ \left [V_{\rm AP} - V_{\rm BP} - V_{\rm AB} +
V_{\rm BB} +
S_{\rm B}^{-1}({\bf q}) \right]
\rho_{\bf q}^{1} \rho_{\bf -q}^{\rm A} \nonumber\\+ \frac{1}{2} \left [ V_{\rm AA} +
S_{\rm A}^{-1}({\bf q}) + V_{\rm BB} +
S_{\rm B}^{-1}({\bf q}) - 2
V_{\rm AB} \right ] |\rho_{\bf q}^{\rm A}|^{2} 
\end{eqnarray}
The above equation takes a greatly simplified form in the matrix notation
\begin{equation}
\beta {\cal H} = \frac{1}{2} \sum_{\bf q} {\vec \rho}_{\bf q}^{T} {\vec S}({\bf
  q})^{-1} {\vec \rho}_{\bf -q}
\end{equation}
where ${\vec \rho} $ is a column vector
\begin{equation}
{\vec \rho}({\bf q}) = {\rho_{\bf q}^{1}\choose \rho_{\bf q}^{\rm A}}
\end{equation}
${\vec \rho}^{T} $ is its transpose and ${\vec S}({\bf q})$ is a second rank square matrix whose elements depend
upon the single particle
structure factors $S_{\rm A}^{0}({\bf q}), S_{\rm B}^{0}({\bf q})$,  $ S_{1}^{0}({\bf q})$ and the
Flory-Huggins interaction parameters $\chi_{\rm AB},  \chi_{\rm BP}$, and
$\chi_{\rm AP}$.
\begin{equation}
{\vec S}^{-1}({\bf q}) = \left(\begin{array}{cc} { S_{1}^{-1}({\bf q}) +
      S_{\rm B}^{-1}({\bf q}) 
- 2 \chi_{\rm BP}}&{ S_{\rm B}^{-1}({\bf q}) - \chi_{\rm AB} - \chi_{\rm BP} + 
\chi_{\rm AP}} \\
{ S_{\rm B}^{-1}({\bf q}) - \chi_{\rm AB} - \chi_{\rm BP} + 
\chi_{\rm AP}}&{ S_{\rm A}^{-1}({\bf q}) + S_{\rm B}^{-1}({\bf q}) - 2
\chi_{\rm AB}}
\end{array}
\right)
\end{equation}
Inversion of this matrix gives the partial structure factors $S_{11}({\bf q})$, $
S_{\rm AA}({\bf q})$ and  $S_{\rm A1}({\bf q})$
\begin{equation}
S_{11}({\bf q}) = \nonumber\\   \frac{S_{\rm A}^{-1}({\bf q}) + S_{\rm B}^{-1}({\bf q}) -  2
  \chi_{\rm AB}}{(S_{1}^{-1}({\bf q}) +  
  S_{\rm B}^{-1}({\bf q}) - 2 \chi_{\rm BP})(S_{\rm A}^{-1}({\bf q}) +
  S_{\rm B}^{-1}({\bf q}) - 2 \chi_{\rm AB})-(S_{\rm B}^{-1}({\bf q}) -
  \chi_{\rm BP} -
  \chi_{\rm AB} +
 \chi_{\rm AP})^{2}}
\end{equation}
\begin{equation}
S_{\rm AA}({\bf q}) = \frac{S_{1}^{-1}({\bf q}) + S_{\rm B}^{-1}({\bf q}) -  2
  \chi_{\rm BP}}{(S_{1}^{-1}({\bf q}) +
  S_{\rm B}^{-1}({\bf q}) - 2 \chi_{\rm BP})(S_{\rm A}^{-1}({\bf q}) +
  S_{\rm B}^{-1}({\bf q}) - 2 \chi_{\rm AB})-(S_{\rm B}^{-1}({\bf q}) -
  \chi_{\rm BP} - \chi_{\rm AB} +
 \chi_{\rm AP})^{2}}
\end{equation}
\begin{eqnarray}
S_{\rm A1}({\bf q}) & =&  S_{\rm 1A}({\bf q}) \nonumber \\  
= & -& 
\frac{S_{\rm B}^{-1}({\bf q})
- \chi_{\rm AB} -\chi_{\rm BP} + \chi_{\rm AP}}{(S_{1}^{-1}({\bf q}) +
  S_{\rm B}^{-1} ({\bf q})- 2 \chi_{\rm BP})(S_{\rm A}^{-1}({\bf q}) +
  S_{\rm B}^{-1}({\bf q}) - 2 \chi_{\rm AB})-(S_{\rm B}^{-1}({\bf q}) -
  \chi_{\rm BP} - \chi_{\rm AB} +
 \chi_{\rm AP})^{2}}
\end{eqnarray}

These results can be used to extract the properties of the mixture both 
in the thermodynamic limit by letting $q = 0$ and at finite $q$ where we can explore
the chain conformation and the spatial distribution of particles within
the medium. We first consider the thermodynamic limit in the following section. 

\subsubsection{The thermodynamic limit ${\bf q}$ = 0}

In this limit $q = 0$ and equations (21) and (22) become
\begin{equation} 
S_{11}^{-1} = \phi_{1}^{-1}+ \phi_{\rm B}^{-1} - 2 \left( \chi_{\rm BP}+ 
\frac{({\phi_{\rm B}^{-1}- \chi_{\rm AB} - \chi_{\rm BP} + \chi_{\rm
      AP}})^{2}}{2(\phi_{\rm A}^{-1} +
  \phi_{\rm B}^{-1} - 2 \chi_{\rm AB})}\right)
\end{equation}
\begin{equation}
S_{\rm AA}^{-1} = \phi_{\rm A}^{-1}+ \phi_{\rm B}^{-1} - 2 \left( \chi_{\rm AB}+ 
\frac{({\phi_{\rm B}^{-1}- \chi_{\rm AB} - \chi_{\rm BP} + \chi_{\rm AP}})^{2}}{2(\phi_{1}^{-1} +
  \phi_{\rm B}^{-1} - 2 \chi_{\rm BP})}\right)
\end{equation}
For simplicity, we remove the argument ${\bf q}$ from the structure factor in the
thermodynamic limit ${\bf q}$ = 0. We consider a mixed solvent in the vicinity of the critical
temperature $T_{c}$ where $\chi_{\rm AB}$ is close to the critical value $\chi_{c}$
and show how a small amount of polymer added to the mixture shifts the
criticality condition. We also examine the way in which the chain responds
to the approach towards solvent criticality condition. 


In the pure solvent mixture when ${\bf q}$ = 0, one has
\begin{equation}
S_{\rm AA}^{-1} = 2 (\chi_{c}-\chi_{\rm AB})
\end{equation}
which can be obtained from equation (25) by letting $\phi_{{1}} = 0$; $\chi_{c}$ is the critical parameter for phase separation which is
function of the solvent composition only
\begin{equation}
\chi_{c} = \frac{1}{2} (\phi_{\rm A}^{-1} + \phi_{\rm B}^{-1})
\end{equation}
If polymer is added to the solution, the partial structure
factor in the forward direction $S_{\rm AA}$ becomes function of the
polymer volume fraction $\phi_{1}$ and can be written as
\begin{equation}
S_{\rm AA}^{-1} = 2 (\chi_{c}'-\chi_{\rm AB}) 
\end{equation}
This result shows that the critical parameter undergoes a shift by
an amount $\Delta \chi_{c}$ 
\begin{equation}
\Delta\chi_{c} = \chi_{c}' - \chi_{c} = - \frac{({\phi_{\rm B}^{-1}- \chi_{\rm AB} -
    \chi_{\rm BP} + \chi_{\rm AP}})^{2}}{2(\phi_{1}^{-1} N_{1}^{-1} +
  \phi_{\rm B}^{-1} - 2 \chi_{\rm BP})}
\end{equation} 
One observes that the critical parameter $\chi_{c}'= \chi_{c}+
\Delta\chi_{c}$ either increases or decreases depending upon the sign of
$\Delta\chi_{c}$ which in turn is fixed by the sign of the denominator in
equation (29). This sign is determined by the relative magnitude of $\chi_{\rm
  BP}$ as compared to
$\frac{1}{2 \phi_{1}N_{1}} + \frac{1}{2 \phi_{\rm B}}$. If  $\chi_{\rm BP}$
the interaction between polymer and solvent B is smaller than $\frac{1}{2
  \phi_{1}N_{1}} + \frac{1}{2 \phi_{\rm B}}$, the quantity $\Delta\chi_{c}$ is negative and the critical parameter decreases
(i.e.$\chi_{c}'< \chi_{c}$). This means that the polymer favors solvent demixing
since a smaller interaction drives the solvents to the limit of
stability.
For example, choosing $N_{1}$= 100, $\phi_{1}$= 0.01 and $\phi_{\rm B}$=0.495, gives
$\frac{1}{2 \phi_{1}N_{1}} + \frac{1}{2 \phi_{\rm B}} = 1.5   $. This means that if
$\chi_{\rm BP}$ is smaller than 1.5, $\Delta\chi_{c}$ is negative and  addition of a small amount of polymer
(approximately 1 percent) induces phase separation of the solvent mixture.
If $\chi_{\rm BP}$ is higher than $1.5$, the system behaves differently since
addition of polymer would result into a positive $\Delta\chi_{c}$ indicating
compatibility enhancement of the solvent mixture. The mixture phase separates
only under a strong repulsion between $\rm A$ and $\rm B$ solvents.  These features
can be 
observed in $figure$ $1$ where $\chi_{c}'$ is 
plotted against $\phi_{1}$ for two
values of $\chi_{\rm BP}$. One observes 
different tendencies, depending upon the
value of the interaction parameter 
$\chi_{\rm BP}$. For the lower value, there is
a slight decrease of $\chi_{c}'$ upon addition of 
polymer. If $\chi_{\rm BP}$ is
as high as 1.5 expressing a large polymer-solvent B 
incompatibility, a small
amount of polymer would result into a large increase 
of $\Delta\chi_{c}$ meaning
that the solvents A and B prefer to remain mixed rather than adsorbing on the
polymer even though the interaction $\chi_{\rm AB}$ chosen in plotting this figure
$(\chi_{\rm AB} = 1)$ is relatively high. This
compatibility enhancement of the mixed solvent due 
to the presence of polymer is quite spectacular since
$\chi_{c}'$ increases by practically an order of magnitude 
upon addition of a small
amount of polymer. Qualitatively, this behavior is consistent with the observation made on PS/(benzene+cyclohexane)
by Varra and Autalik \cite{Vavra:97.1}. These authors found that a strong interactions between PS
and non solvents (cyclohexane) could lead to a reduced solvent-solvent interaction
and hence a compatibility enhancement of the mixture.
Furthermore, one finds that the shift $\chi_{c}' - \chi_{c} =
\Delta\chi_{c}$ is only moderately sensitive to the interaction asymmetry
between the polymer and the solvents A and B which is expressed by the parameter $\epsilon =
\chi_{\rm BP}-\chi_{\rm AP}$. The shift in the critical parameter $\Delta\chi_{c}$
is strong upon addition of polymer even for small values of $\epsilon$. The
interaction asymmetry parameter plays a
more crucial role in connection with the polymer conformation near the solvent
criticality condition as we shall see in the following section.


The effective potential $V_{0}$ of equation (16) is
directly related to the
classical excluded volume parameter acting on a polymer in the infinite dilute
limit. This is made clear by writing the partial structure factor $S_{11}({\bf
  q})$ in the reciprocal form
similar to Zimm's equation
\begin{equation}
S_{11}^{-1}({\bf q}) = S_{1}^{-1}({\bf q}) + V_{0} 
\end{equation}
Letting $\Delta\chi_{0} = \chi_{c} - \chi_{\rm AB}$ one obtains
\begin{equation}
V_{0} = \frac{1}{\phi_{\rm B}}- 2 \chi_{\rm BP} -
\frac{(\Delta\chi_{0} + \epsilon)^{2}}{2 \Delta\chi_{0}} 
\end{equation}
For an upper critical solution temperature mixture, the interaction parameter
$\chi_{\rm AB}$ increases approaching $\chi_{c}$. In this case, the polymer
reacts in different ways depending essentially on whether $\epsilon$ is zero
or not.
This is a clear illustration of the importance of interaction asymmetry  
between polymer and solvents in determining the thermodynamic behavior of the
solution. As the solvent mixture approaches the criticality condition ($\Delta\chi_{0}\to 0$), a striking difference exists between the
case where the solvents present the same interaction with the polymer ($\epsilon
= 0$) and the case where the two solvents interact differently with the
polymer ($\epsilon \neq 0$). Equation (31) shows that in the case of a
symmetric interaction $(\epsilon = 0)$, $V_{0}$ decreases linearly with
$\Delta\chi_{0}$. However, in the non symmetric interaction problem  $(\epsilon \neq 0)$, the
third term in the RHS of equation (31) increases rapidly as
$\Delta\chi_{0}\to 0$ and $V_{0}$ undergoes a rapid decrease. Following this sharp
decrease, the effective potential vanishes and may even become negative as
$\chi_{\rm AB}$  approaches $\chi_{c}$. The chain undergoes a phase transition from a
swollen state to unperturbed dimensions or even a collapsed configuration. This is illustrated in $figure$ $ 2$ where $ V_{0}$ is
represented as a function of $\Delta\chi_{0}$ for $\epsilon = 0$ and 0.2. The
upper curve in this figure corresponds to the case where the
interaction between polymer and solvents A and B are the same and $\epsilon =
0$. It shows that $V_{0}$
increases relatively smoothly when  $\Delta\chi_{0} \to 0$. 

One could also introduce an effective polymer-solvent
interaction parameter $\chi_{0}$ via 
\begin{equation}
V_{0} = \frac{1}{\phi_{\rm B}}- 2 \chi_{0}  
\end{equation}
With
\begin{equation}
\chi_{0} =  \chi_{\rm BP}+ 
\frac{(\Delta\chi_{0} + \epsilon)^{2}}{4 \Delta\chi_{0}} 
\end{equation}
$Figure$ $ 3$ represents the variation of the effective polymer-solvent
interaction parameter $\chi_{0}$ as a function of  $\Delta\chi_{0}$
for the same values of $\epsilon$ as in $figure$ $ 2$ and shows similar
tendencies. For $\epsilon =0 $, $\chi_{0}$
decreases slightly when $\Delta\chi_{0}\to 0$, while
for $\epsilon = 0.2$, there is an inversion in the variation of $\chi_{0}$
expressed by a turn up and a sharp increase as the mixed solvent approaches
criticality indicating a strong
polymer-solvent repulsion. 

\subsubsection{The small ${\bf q}$ limit}

As a first approximation, we consider that $q^{-1}\gg 1$ which means that by
scattering radiation,one observes the solvent particles as point like and the $q$-
dependence comes entirely from the form factor of the polymer only. This approximation could be easily
relaxed if one wants to include intramolecular interferences within the solvent
particles themselves and explore by radiation scattering the internal
structure of chains. Moreover, we will assume that the form factor of the chain could
be approximated by  
\begin{equation}
P(q) = \frac{1}{1 + \frac{q^{2}R_{\rm g}^{2}}{2}}
\end{equation}
Following these assumptions, equation (17) can be written in the form 
\begin{equation}
S_{11}(q) = \frac{S_{11}(q=0)}{1 + q^{2}\xi_{0}^{2}}
\end{equation}
Where the correlation length $\xi_{0}$ represents the distance over which the
polymer configuration is sensitive to solvent fluctuations. It depends upon the polymer radius of
gyration $R_{\rm g}$ and the excluded volume interaction $V_{0}$ according to the relationship  
\begin{equation}
\xi _{0}^{2} = \frac{R_{\rm g}^{2}}{2(1 + V_{0}\phi_{1}N_{1})}
\end{equation}
Figure(4) represents the variation of $\xi_{0}/R_{\rm g}$ as a function of
$\Delta\chi_{0}$ for $\epsilon$ = 0 and 0.2. For $\epsilon = 0$,
$\xi_{0}/R_{\rm g}$ decreases
continuously when $\Delta\chi_{0}\to 0$ while for $\epsilon =
0.2$, $\xi_{0}/R_{\rm g}$ increases suddenly when $\Delta\chi_{0}$ approaches zero. One can
visualize the polymer as a chain of blobs of size $\xi_{0}$. Over distances
below the blob size, the chain is sensitive to solvent fluctuations and its
size reduces as a result of increasing effective interaction $\chi_{0}$. Over distances exceeding the blob size, the chain does not feel
the effects of solvent fluctuations and shows a comparatively swollen conformation. As
the solvent approaches criticality when $\Delta\chi_{0}\to 0$, the correlation
length $\xi_{0}$ becomes comparable or larger than $R_{\rm g}$ implying that the
entire chain is
subject to solvent fluctuations and undergoes a phase transition from a
swollen to a more compact conformation. For $\epsilon = 0$ and
$\chi_{BP} = \chi_{AP} = 0.5$, the chain feels theta solvent conditions
even though $\Delta\chi_{0}\to 0$ and solvents demixing
conditions are approached. 

The blob picture for a polymer in mixed solvents was first
suggested by  Brochard and de Gennes \cite{Brochard:80.1} who examined the polymer
conformation in a mixture of two good solvents, when the polymer affinity
differs substantially for the two solvents. They were the first to predict
that the polymer adopts a collapsed configuration near $T_{c}$ even
it would be swollen in either of the pure solvents under similar conditions.

\section{Many chains in mixed solvents}

\subsection{The effective potential of a test chain}

 In this section, we discuss the problem of many chains by extending the
 Edwards Hamiltonian formalism to strong solutions of polymers and mixed solvents. The effective potential acting on a test chain  together with
partial structure factors are calculated as a function of polymer
concentration and mixed solvent properties. The Hamiltonian formalism is
extended by taking into account the effects of the many chains present in the medium. This gives the following result for the
partition function including summations over all chains designated by the
letters $\alpha$ and $\beta$   

\begin{eqnarray}
{ \cal Z}= \int \prod {\cal D}{{\bf R}_{\alpha}(s)} \int \prod
d\vec{{\bf r}_i^A} \prod d\vec{{\bf r}_i^B} \exp [  - \frac{3}{2 l^{2}}\sum_{\alpha}\int_0^N
\left( \frac {\partial {\bf R}_{\alpha}(s)}{\partial S} \right)^{2}ds \nonumber\\ 
-\frac{1}{2} V_{\rm PP} \sum_{\alpha \beta} \int_0^N\int_0^N
\delta({\bf R}_{\alpha}(s) - {\bf R}_{\beta}(s')) ds ds' \nonumber\\ -
V_{\rm AP} \sum_{\alpha} \int_0^N \sum_{i} \delta({\bf R}_{\alpha}(s) - {\bf r}_i^A) ds
- V_{\rm BP} \sum_{\alpha} \int_0^N \sum_{i} \delta({\bf R}_{\alpha}(s) - {\bf
  r}_i^B)ds \nonumber\\  -
\frac{1}{2} V_{\rm AA} \sum_{ij} ({\bf r}_i^A - {\bf r}_j^A) - \frac{1}{2}
V_{\rm BB}
 \sum_{ij} \delta ({\bf r}_i^B - {\bf r}_j^B) - V_{\rm AB} \sum_{ij} \delta
({\bf r}_i^A-{\bf r}_j^B) ]  
\end{eqnarray}
The subscript $\alpha$ and $\beta$ run over all chains in the solution
except a test chain designated by the subscript $1$ (meaning
single). Transforming space coordinates into collective density variables
leads to the partition function 
\begin{eqnarray}
{\cal Z} = \int {\cal D}{{\bf R}_{1}(s)} \int \prod \delta {\bf \rho}_{\bf q}^{\rm A} \prod
\delta {\bf \rho}_{\bf q}^{\rm B} \exp{- \frac {3}{2l^{2}}} \int_{0}^{N} \left ( \frac {\partial
  {\bf R}_{1}(s)}{\partial
  s}\right) ^{2} ds - \frac{1}{2} V_{\rm PP} | {\bf \rho}^{1}|^{2} \nonumber\\
 - V_{\rm PP}
\sum_{\bf q} |{\bf \rho}_{\bf q}^{\rm P}||{\bf \rho}_{\bf q}^{1}| - \frac{1}{2}
\sum _{\bf q} ( V_{\rm PP} +
S_{\rm P}^{-1}({\bf q}) )|{\bf \rho}_{\bf q}^{\rm P}|^{2} \nonumber\\ 
 - V_{\rm AP} \sum_{\bf q} {\bf \rho}_{\bf q}^{1} {\bf \rho}_{\bf -q}^{\rm A} -
V_{\rm AP} \sum_{\bf q} {\bf \rho}_{\bf q}^{\rm P} {\bf \rho}_{\bf- q}^{\rm A}
- V_{\rm BP} \sum_{\bf q} {\bf \rho}_{\bf q}^{1} {\bf \rho}_{\bf -q}^{\rm B} -
V_{\rm BP} \sum_{\bf q}
|{\bf \rho}_{\bf q}^{\rm P}|
|{\bf \rho}_{\bf q}^{\rm B}| \nonumber\\ - \frac{1}{2} \sum_{\bf q} ( V_{\rm AA} + S_{\rm A}^{-1}({\bf q}) )|{\bf \rho}_{\bf q}^{\rm A}|^{2} - \frac{1}{2} 
\sum _{\bf q}(V_{\rm BB} + S_{\rm B}^{-1}({\bf q}) )|{\bf \rho}_{\bf q}^{\rm B}|^{2} -
\sum _{\bf q} V_{\rm AB}
{\bf \rho}_{\bf q}^{\rm A} {\bf \rho}_{\bf -q}^{\rm B} 
\end{eqnarray}
Where  ${\bf \rho}_{\bf q}^{\rm A}$ and ${\bf
  \rho}_{\bf q}^{\rm B}$, ${\bf \rho}_{\bf q}^{\rm P}$ and ${\bf \rho}_{\bf
  q}^{1}$ are the collectives coordinates for the solvents $A$ and $B$, the
polymer $P$ and the test chain $1$, respectively. In order to identify the
effective potential acting on a test chain together with the partial scattering functions for the multicomponent mixture under investigation, it is
convenient to write the partition function in matrix form
\begin{eqnarray}
 {\cal Z} = \int {\cal D}{{\bf R}_{1}(s)} \int \prod \delta {\bf
   \vec{\rho}}_{\bf q} \exp{[- \frac
   {3}{2l^{2}} \int_{0}^{N} \left( \frac {\partial {\bf R}_{1}(s)}{\partial 
  s}\right )^{2} ds  - \frac{1}{2} V_{\rm PP} \sum |{\bf \rho}_{\bf
q}^{1}|^{2}}] \nonumber\\
\exp [{ - \sum 2  \vec{{\bf V}}({\bf q})  \vec{{\bf \rho}}_{\bf q} - \sum 
  \vec{{\bf \rho}}_{\bf q}^{-1}  \vec{{\bf M}}({\bf q})  \vec{{\bf \rho}}_{\bf
    -q}}]
\end{eqnarray}
where $\vec{\bf \rho}_{\bf q}$ and $\vec {\bf V}({\bf q})$ represent the column
vectors 
\begin{equation}
\vec{\bf \rho}_{\bf q} = \left(\begin{array}{cc} {\bf \rho}_{\bf q}^{\rm A}\\ {\bf
      \rho}_{\bf q}^{\rm B}\\ {\bf \rho}_{\bf q}^{\rm P}
  \end{array}\right) 
\end{equation}
\begin{equation}
{\bf \vec V}({\bf q}) = \left(\begin{array} {cc} \frac{1}{2} V_{\rm AP}
    \int_{0}^{N} \exp (- \rm i{\bf q
  R}(s) ds) \\\frac{1}{2} V_{\rm BP} \int_{0}^{N} \exp (- \rm i{\bf q
  R}(s) ds) \\ \frac{1}{2} V_{\rm PP} \int_{0}^{N} \exp (- \rm i{\bf q
  R}(s) ds) \end{array}\right )  
\end{equation}
and $ {\bf \vec M}({\bf q})$ is a square three-by-three matrix 
\begin{equation}
{\bf \vec M}({\bf q}) = \left(\begin{array}{ccc}
    \frac{1}{2}(V_{\rm AA}+\frac{1}{S_{\rm A}({\bf q})}) &
    \frac{1}{2} V_{\rm AB} & \frac{1}{2} V_{\rm AP} 
\\ \frac{1}{2}V_{\rm AB} & \frac{1}{2}(V_{\rm BB} + \frac{1}{S_{\rm B}({\bf q})}) &
\frac{1}{2} V_{\rm BP} \\ \frac{1}{2} V_{\rm AP} & \frac{1}{2} V_{\rm BP} &  \frac{1}{2}
(V_{\rm PP} + \frac{1}{S_{\rm P}({\bf q})}) \end{array}\right )
\end{equation}

Integrating over the collective variables $\rho_{\bf q}$, one obtains the
partition function as follows

\begin{equation} 
{\cal Z} = \int {\cal D}{{\bf R}_{1}(s)}  \exp \left [{{- \frac
   {3}{2l^{2}} \int_{0}^{N} \left( \frac {\partial {\bf R}(s)}{\partial 
  s}\right)^{2}}  ds - \frac{1}{2} V_{\rm PP} \sum_{\bf q} |{\bf \rho_{q}}^{1}|^{2} +
\sum_{\bf q} {\bf \vec V_{q}}^{-1} {\bf \vec
M_{q}}^{-1} {\bf \vec V_{q}} }\right ] 
\end{equation}

Recalling that ${\cal Z} = \int {\cal D}R_{1}(s) \exp - \beta {\cal H}$
yields
\begin{equation}
\beta {\cal H} = \frac{3}{2l^2} \int_0^N \left (\frac{\partial {\bf R}(s)}{\partial
  s}\right )^{2}ds + \sum_{\bf q} V({\bf q}) \int_0^N\int_0^N ds ds' \exp \left (- \rm i{\bf
  q} ({\bf R}(s)-{\bf R}(s'))\right )
\end{equation}
where one can identify the effective potential $V_{\bf q}$ acting on a test
chain. If the volumes occupied
by monomers and solvent molecules are the same, then $V_{\rm PP}$, $V_{\rm AA}$ and
$V_{\rm BB}$ are approximately represented by the same
quantity denoted $V_{\infty}$ and one can introduce
the Flory-Huggins interaction parameters using $V_{\infty}$ = $ V_{ij}$ -
$\chi_{ij}$ where the subscripts i and j run over A, B and P. All polymer
chains in the medium have the same 
interaction parameters and no distinction exists between the test chain 1 and the
others. The effective potential $V({\bf q})$ is not only due to the solvents but
also to all the chains present in the medium. For incompressible mixtures, the
potential $V_{\infty}$ goes to infinity and one obtains 

\begin{equation}
V({\bf q}) = \frac{S_{\rm A}^{-1}({\bf q})  S_{\rm B}^{-1}({\bf q}) + D - 2
  \chi_{\rm AP} S_{\rm B}^
  {-1}({\bf q})  - 2 \chi_{\rm BP}
  S_{\rm A}^{-1}({\bf q}) } {S_{\rm A}^{-1}({\bf q}) + S _{\rm B}^{-1}({\bf q}) - 2
  \chi_{\rm AB} + S_{\rm P} ({\bf q}) (S_{\rm A}^{-1}({\bf q})
  S_{\rm B} ^{-1}({\bf q}) + {\rm D} - 2 \chi_{\rm AP} S_{\rm B}^{-1}({\bf q})
  - 2 \chi_{\rm BP}
  S_{\rm A}^{-1}({\bf q}))}
\end{equation}
Here ${\rm D}$ is defined in equation (15). It is interesting to note
that the effective potentials acting on chains in infinitely dilute solutions
$V_{0}({\bf q})$ and in strong solutions $V({\bf q})$ are related via 
\begin{equation}
{ V}^{-1}({\bf q}) = {V}_{0}^{-1}({\bf q}) + {S}_{\rm P}({\bf q})
\end{equation}
This result is characteristic of the series expansion in the Ornstein-Zernike model\cite{Vilgisben:91.1}. It suggests that one can write the total  potential $V({\bf q})$
as  
an infinite series of interaction terms. The order of these terms  depends on the
number of intermediate 
chains involved in the interaction (see $figure$ $ 5a$). Each diagram corresponds to a
particular term in the series expansion
\begin{equation}
V ({\bf q})= V_{0}({\bf q}) - V_{0}({\bf q}) S_{\rm P}({\bf q}) V_{0}({\bf q}) +
V_{0}({\bf q}) S_{\rm P}({\bf q}) V_{0}({\bf q})  S_{\rm P}({\bf q}) V_{0}({\bf q}) + ...
\end{equation}
These series can be summed up in a similar way as in the Ornstein-Zernike 
expansion ( see figure 5b). The summation gives
\begin{equation}
V({\bf q}) = V_{0}({\bf q}) - V_{0}({\bf q}) S_{\rm P}({\bf q}) V({\bf q})
\end{equation}
which is equivalent to
\begin{equation}
V({\bf q}) = V_{0}({\bf q}) \left [ 1 + V_{0}({\bf q}) S_{\rm P}({\bf q})\right ]^{-1}
\end{equation}

\subsection{The Edwards screening}
For completeness, we remark that the Edwards screening can be recovered.
The effective potential for many chains in equation (47) can also
be written in the following form
\begin{equation}
V({\bf q}) = V_{0}({\bf q}) - \frac{V_{0}^{2}({\bf q}) }{S_{\rm P}^{-1}({\bf q}) +
  V_{0}({\bf q}) }
\end{equation}
Using Edwards prescription for the form factor of  Gaussian chains
\begin{equation}
S_{\rm P} = \frac{\phi_{\rm P} N_{\rm P}}{1 + \frac{1}{2}q^{2}R_{\rm g}^{2}}
\end{equation}
where the radius of gyration is related to the degree of polymerization $N$
and segment length $l$ by the known result $R_{\rm g} = l \sqrt{N/6}$. Substituting  equation 51 into 50 yields
\begin{equation}
 \frac{V}{V_{0}} = \frac{q^{2}\xi^{2}}{1 + q^{2}\xi^{2}}
\end{equation}
Where $\xi$ is the Edwards screening length
\begin{equation}
\xi = \frac{l}{\sqrt{12 V_{0} \phi_{\rm P}}}
\end{equation}
In the infinite dilute limit, $\phi_{\rm P}$ goes to 
zero and the screening length
$\xi$ tends to infinity. As one would expect, the effective potential $V$
tends to $V_{0}$ in this limit.
Figure 6 represents the ratio $V/V_{0}$ versus $q \xi$ as given by
equation 52. The ratio $V/V_{0}$ is small for $q\xi$ less than $1$ 
indicating screening of the potential field acting on a test
chain. When $q\xi$ increases the ratio $V / V_{0}$ tends to 1 and the
effective potential is practically the same as in infinite dilute
solutions. Locally at large $q$, the test chain behaves as if it were 
isolated and
is essentially not perturbed by the presence of other chains in the
medium. It feels only the fluctuations of the solvent mixture in its
immediate vicinity.

This analysis of screening presented here and following the Edwards
prescription is only qualitative. A more precise treatment is needed if one
wants to have a quantitative evaluation of the screening effects. Such an
improvement can be made using other methods such as
renormalization group theory methods or field theoretical tools 
\cite{DesCloizeaux:90.1,Freed:87.1,Vilgms:98.1}.

\subsection{The scattering functions}

The scattering matrix can be readily extracted from the partition function in
the matrix notation 
\begin{equation}
{\bf S}^{-1}({\bf q}) = \left( \begin{array} {ccc} S_{1}^{-1}({\bf q}) +
    S_{\rm B}^{-1}({\bf q}) - 2 \chi_{\rm BP} & S_{\rm B}^{-1}({\bf q}) - 2
    \chi_{\rm BP} &
    S_{\rm B}^{-1}({\bf q}) - \chi_{\rm AB} - \chi_{\rm BP} + \chi_{\rm AP} \\
    S_{\rm B}^{-1}({\bf q}) - 2 \chi_{\rm BP}
    & S_{\rm P}^{-1}({\bf q}) + S_{\rm B}^{-1}({\bf q}) - 2 \chi_{\rm BP} &
    S_{\rm B}^{-1}({\bf q}) - \chi_{\rm AB} -
    \chi_{\rm BP} + \chi_{\rm AP} \\  S_{\rm B}^{-1}({\bf q}) - \chi_{\rm AB}
    - \chi_{\rm BP} + \chi_{\rm AP} &
    S_{\rm B}^{-1}({\bf q}) - \chi_{\rm AB} - \chi_{\rm BP} + \chi_{\rm AP} &
    S_{\rm A}^{-1}({\bf q}) + S_{\rm B}^{-1}({\bf q}) -
    2 \chi_{\rm AB} \end{array} \right)
\end{equation}

The diagonal elements of the scattering matrix $S({\bf q})$ define the
contribution to the scattering signal due to different constituents in the mixture. Three quantities are particularly
relevant for our purpose in the present work. Considering the solvent
$B$ as a background medium, one can obtain the diagonal elements of equation
(54) denoted  $S_{11}({\bf q})$, $ S_{\rm PP}({\bf q})$ and $S_{\rm AA}({\bf q})$. The properties of these partial structure
factors are briefly discussed below.

\subsubsection{Structure factor of a test chain $S_{11}({\bf q})$}
The first element of the matrix in equation (54) gives the scattering function
from the test chain
\begin{equation}
S_{11}^{-1}({\bf q}) = S_{1}^{-1} ({\bf q})+ V_{1}({\bf q}) 
\end{equation}
where $V_{1}(\bf q)$ is identical to $V({\bf q})$ given in equation (46). The subscript $1$ is introduced to describe a generalized excluded volume
experienced by a test chain labeled $1$ in
the strong solution. Discussion of the screening presented above could be
applied here invoking the excluded volume interaction experienced by the test
chain $1$.
It is worthwhile to note that if the mixed solvent satisfies the
criticality condition in the absence of polymer, then $\frac{1}{S_{\rm A}} +
\frac{1}{S_{\rm B}} = 2 \chi_{AB}$, and equations (55) and (45) give
\begin{equation}
\frac{1}{S_{11}} = \frac{1}{S_{\rm P}} + \frac{1}{S_{1}}
\end{equation}
This result corresponds to the scattering function, one obtains from a solution of
identical chains under theta conditions where excluded volume
interactions are completely screened out. It is consistent with the variation
of $V_{0}$ versus $\Delta\chi_{0}$ which shows that the apparent excluded
volume tends to 0 as $\Delta\chi_{0} \to 0$ ($figure$ $2$)  
 
\subsubsection{Structure factor of the polymer $S_{\rm PP}(\bf q)$}

The second diagonal term of the scattering matrix $S({\bf q})$ represents the contribution from
the polymer and found as
\begin{equation}
S_{\rm PP}^{-1}({\bf q}) = S_{\rm P}^{-1}({\bf q}) + V_{\rm P}(\bf q) 
\end{equation}
Where $V_{\rm P}(\bf q)$ has a similar expression as $V_{1}(\bf q)$ except that
$S_{\rm P}(\bf q)$ and $S_{1}(\bf q)$ are interchanged.
If the mixed solvent approaches criticality, the signal from the p-chains become
identical to the signal from the test chain $1$ and hence the same as in
equation (56). The chains are under theta condition as indicated earlier.

\subsubsection{Structure factor of the mixed solvent $S_{\rm AA}(q)$}

The third diagonal element of the matrix in equation (54) gives the partial
structure factor of the mixed solvent
\begin{equation}
S_{\rm AA}^{-1}({\bf q}) = S_{\rm A}^{-1}({\bf q}) + S_{\rm B}^{-1}({\bf q}) - 2
\chi_{\rm AB}-  
 \frac{2 (S_{\rm B}^{-1}({\bf q}) -  \chi_{\rm AB} -\chi_{\rm BP} +
   \chi_{\rm AP})^{2}(S^{-1}_{1}({\bf q}) + S^{-1}_{\rm P}({\bf q}))}{ (S_{\rm P}^{-1}({\bf
     q}) + S_{1}^{-1}({\bf q}) )(S^{-1}_{\rm B}({\bf q}) - 2 \chi_{\rm BP}) +
   S_{1}^{-1}({\bf q}) S^{-1}_{\rm P}({\bf q})}
\end{equation}

In the infinite dilute limit $S_{\rm P} = 0$ and one recovers the former result
given in
 equation (22).
Considering the $q$ = 0 limit, one can make similar remarks as those
following equation (29), except that the concentration range
of polymer is not limited to infinite dilute
solutions and the polymer concentration range is extended to higher values
in investigating the variation of the interaction parameter
$\chi'_{c}$ as a function of polymer concentration $\phi_{\rm P}$.

\section{Conclusions}
In this paper, we present some theoretical aspects of polymers in mixed
solvents using the Edwards Hamiltonian formalism. Discussions are confined to
thermodynamic and structural properties and both the single
and the many chain problems are considered. If one solvent species presents a
strong repulsion towards the polymer, addition of a small amount of polymer leads to compatibility enhancement of the solvent mixture. The
critical parameter at which the solvent mixture phase separates increases
dramatically. The effective potential acting on the polymer in the medium
depends enormously on the asymmetry in the affinity of the polymer with respect
to both solvent
species. If this asymmetry is large, as the mixed solvent approaches
the criticality condition, the effective interaction polymer-solvent increases
dramatically. The polymer undergoes a phase transition from a
swollen configuration to unperturbed dimensions and eventually to a collapsed configuration as the mixed solvent approaches
criticality. When the correlation length for solvent fluctuations is of the
order or higher than the radius of gyration of polymer $R_{\rm g}$, the chain
conformation becomes characteristic of a polymer in bad solvents and the
effective excluded volume parameter vanishes and may even become negative. 

In the many chain problem of strong
solutions, the effective potential acting on a test chain is calculated using
the same formalism based on the Edwards Hamiltonian. This potential is related to the bare
effective potential in the infinite dilute limit via a series expansion which
is reminiscent of the Ornstein-Zernike model. This series expansion is
represented graphically via a diagrammatic method where the successive
diagrams represent interactions involving increasing numbers of
intermediate chains. If the solvent mixture is critical, interesting results are
obtained for the structure factor of the polymer and a test
chain. If the solvent mixture is near criticality, all excluded volume
interactions are screened out and chains in the medium feel theta solvent conditions.
  
This investigation is based upon the Edwards Hamiltonian formalism and gives a
qualitative picture of the thermodynamic behavior and structural properties of
polymers in mixed solvents. Near critical regions where the fluctuations are
strong, one has to rely on an appropriate model to achieve a more rigorous
quantitative evaluation of the effects of critical fluctuations on the
thermodynamic and the scattering properties of polymers in mixed solvents. This
could be done within the framework of field theoretical or
renormalization group theory calculations.

\begin{center} 
$Acknowledgements$
\end{center}
This work has been accomplished during a stay of the first
author (A. Negadi) at the MPI-P in Mainz which was made possible by
fellowships from the DAAD (July-September 1998) and from the MPI-P
(October-December 1998). A Negadi and M. Benmouna thank Professor K. Kremer
for his kind invitation and hospitality.


\pagebreak
\begin{center}
Figure Captions
\end{center}
{\bf Figure 1:}\\
Variation of $\chi'_{c}$ the critical interaction parameter between solvents
$A$ and $B$ with the polymer volume fraction $\phi_{1}$ for two values of
$\chi_{BP}$ and $\chi_{\rm AB}$ = 1, $N=100$, $\phi_{\rm B}=0.495$
$\epsilon = \chi_{\rm AP} - \chi_{\rm BP}$ = 0.2 (see
equation (29)). \\
\\
{\bf Figure 2:}\\
Variation of $V_{0}$ the effective excluded volume parameter  with the
parameter $\Delta\chi_{0}$ for two values of
$\epsilon$ and $\chi_{\rm BP}$ =0.5, (see
equation (31), the other values are chosen to be the same as above)\\
\\
{\bf Figure 3:}\\
Variation of $\chi_{0}$ the effective interaction parameter  with the
parameter $\Delta\chi_{0}$ for two values of
$\epsilon$ and $\chi_{\rm BP}$ = 0.5, (see
equation 33)\\
\\
{\bf Figure 4:}
Variation of $\xi_{0}/R_{\rm g}$ the normalized correlation length for solvent fluctuations with the
parameter $\Delta\chi_{0}$ for two values of
$\epsilon$ and $\chi_{\rm BP}$ = 0.5, (see
equation (36))
In plotting these figures, we have used $N = 100$, $\phi_{\rm A}$ = $\phi_{\rm B}$ and
$\chi_{AB}$ = 1.\\
\\
{\bf Figure 5:}
Diagrammatic representation of the relationship between the effective
potentials in the many chain ($V$) and in the single chain problems ($V_{0}$)
a) Representation of the infinite series of equation (47)
b) Representation of equation (48).
Vertical thick lines represent the main interacting chain.
Vertical dashed lines represent intermediate chains of interaction.
Horizontal dashed wavy line is an interaction $V_{0}$Horizontal thick wavy
line is an interaction $V$.\\
\\
{\bf Figure 6:}
 Variation of the ratio $V/V_{0}$ as a function of $q\xi$ according to
equation (52) where $\xi$ is the Edwards screening length.

\pagebreak
\begin{figure}
\epsfxsize=15cm
\epsffile{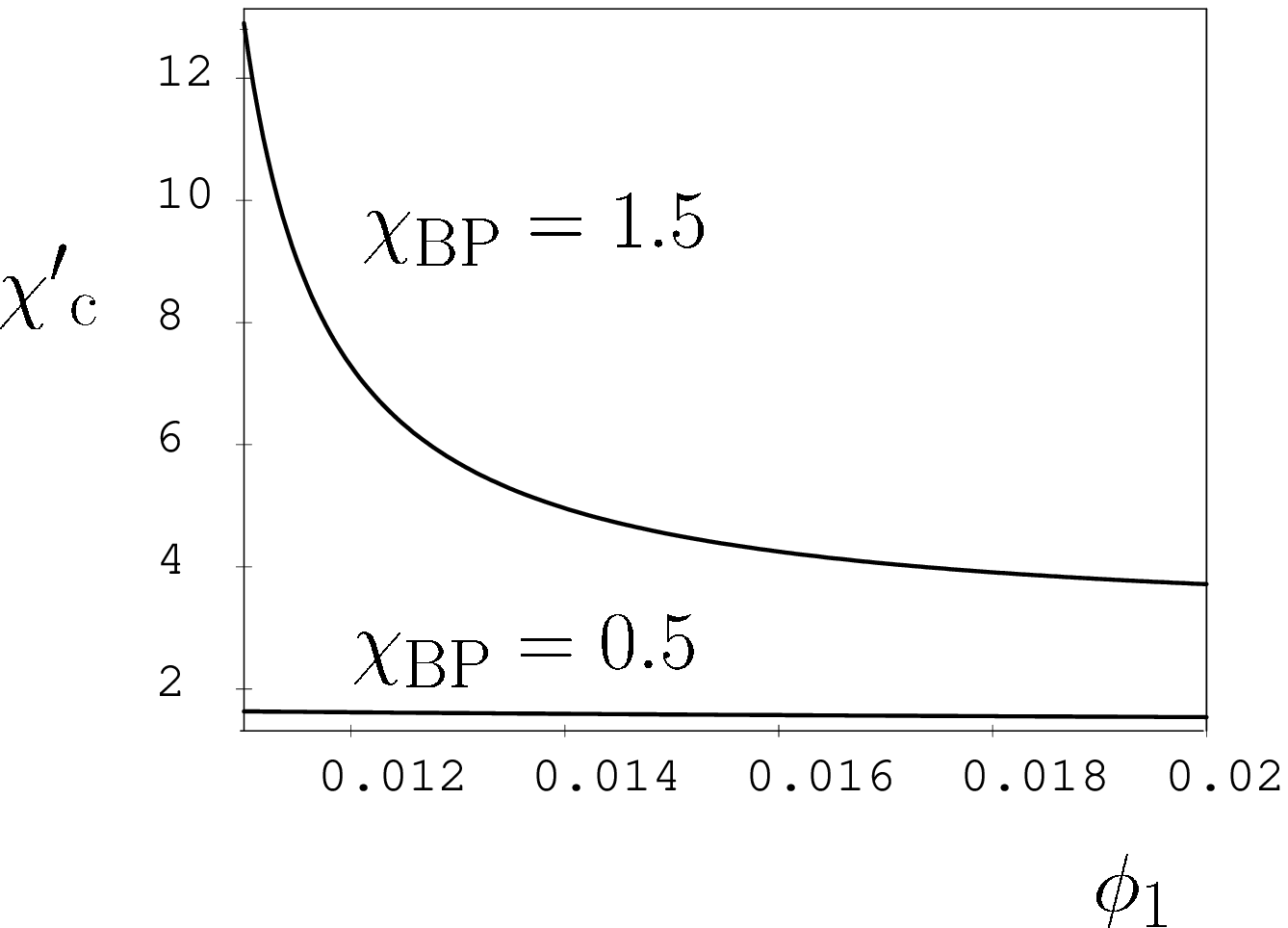}
\end{figure}

\pagebreak
\begin{figure}
\epsfxsize=15cm
\epsffile{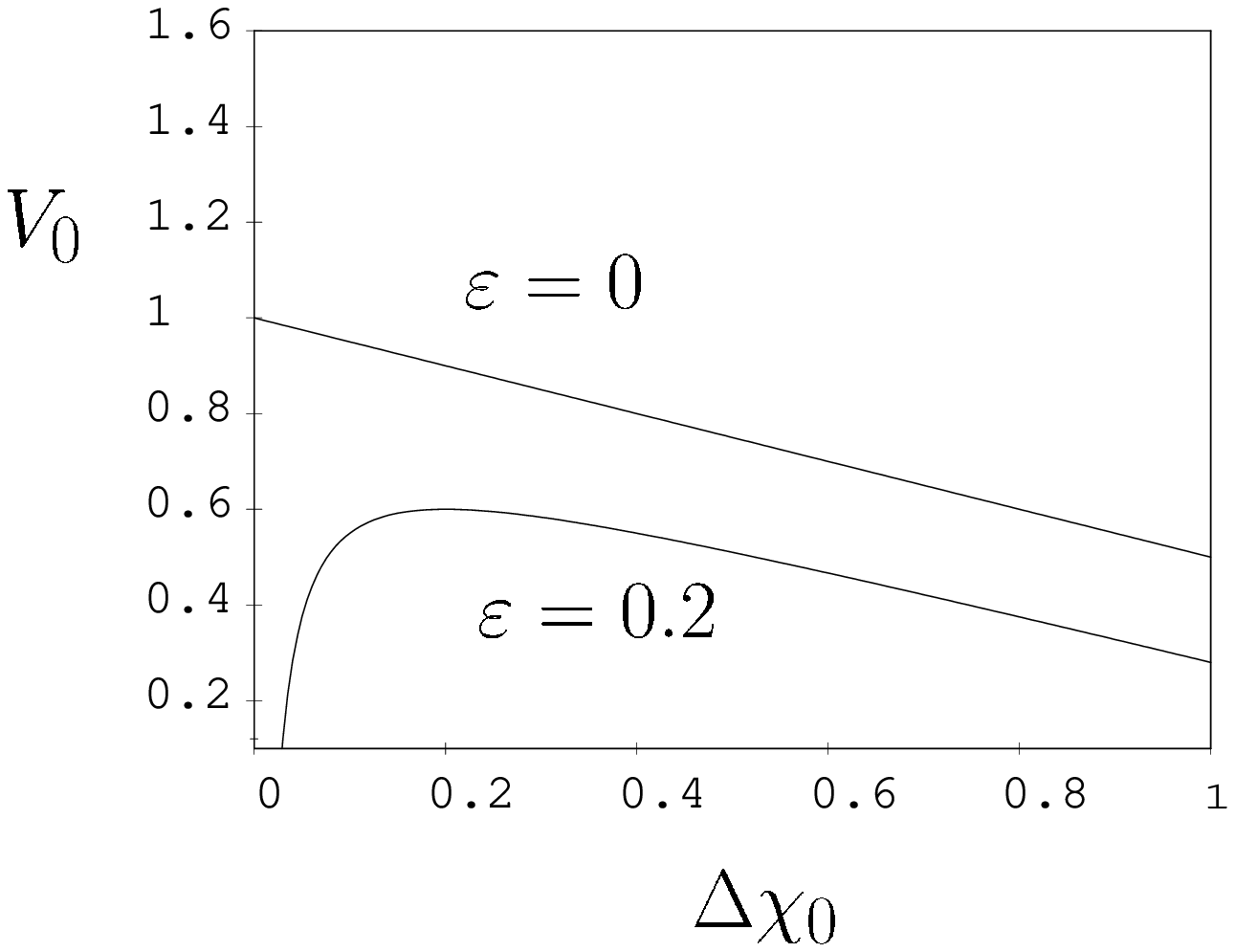}
\end{figure}
\pagebreak
\begin{figure}
\epsfxsize=15cm
\epsffile{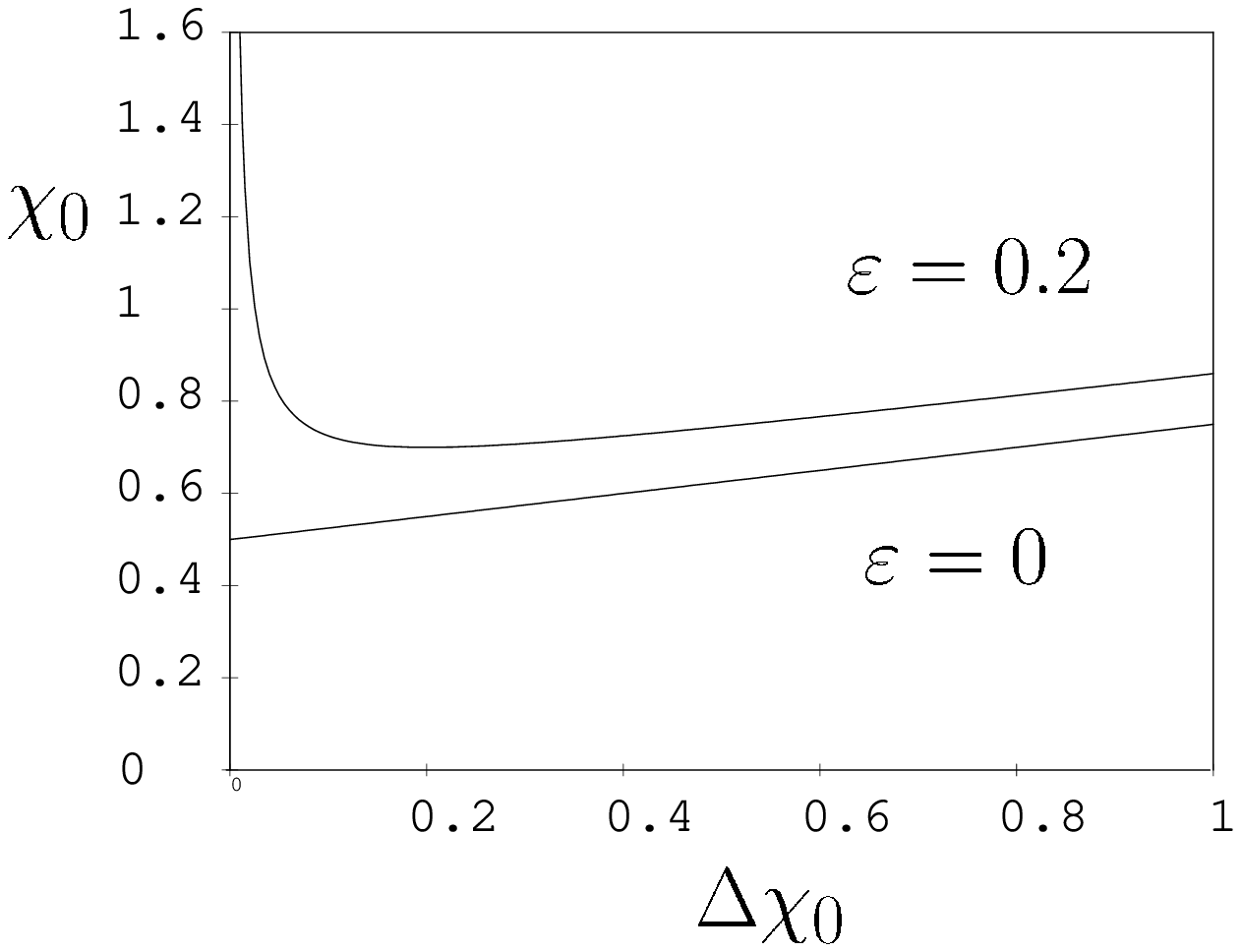}
\end{figure}

\pagebreak
\begin{figure}
\epsfxsize=15cm
\epsffile{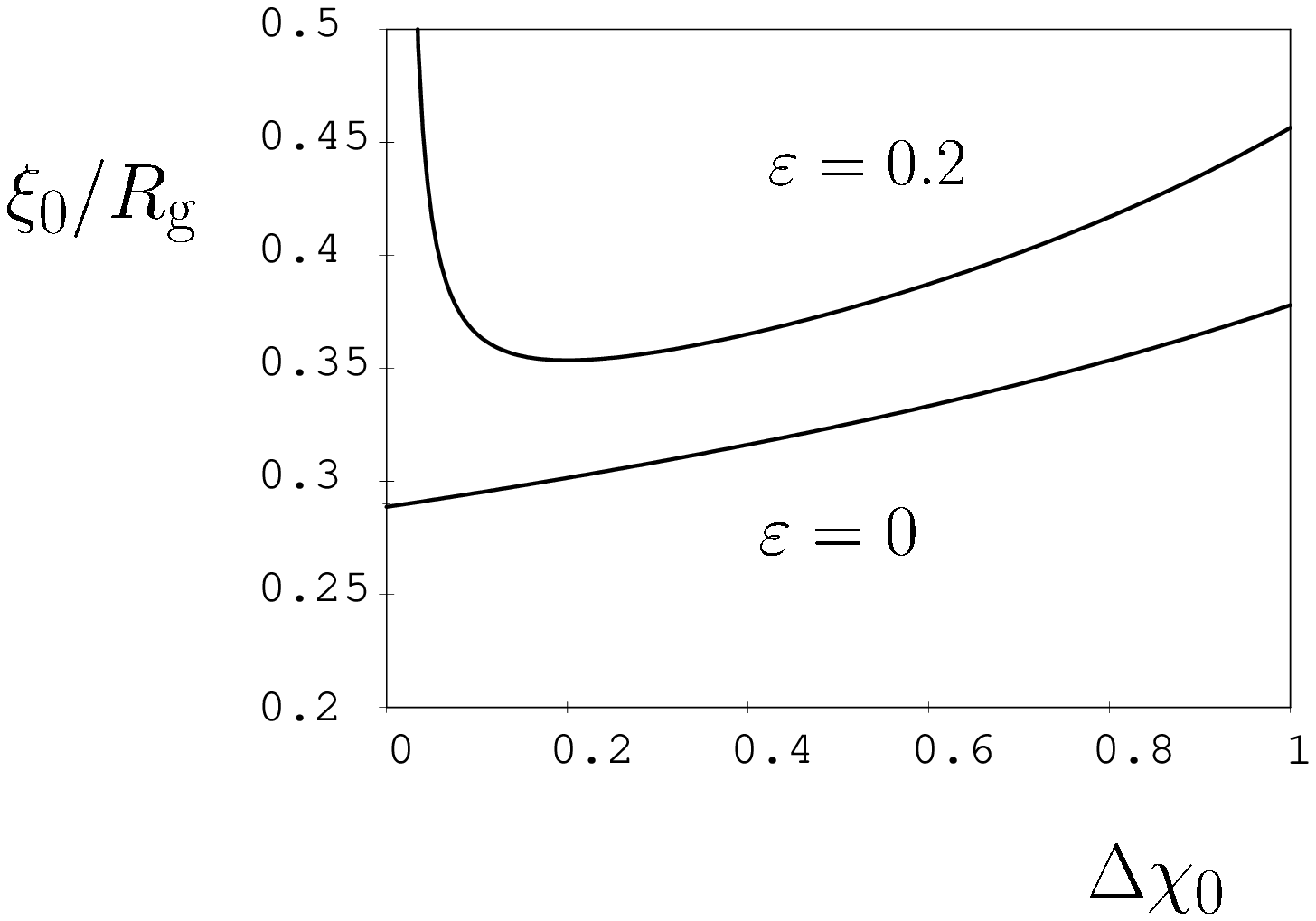}
\end{figure}

\pagebreak
\begin{figure}
\epsfxsize=15cm
\epsffile{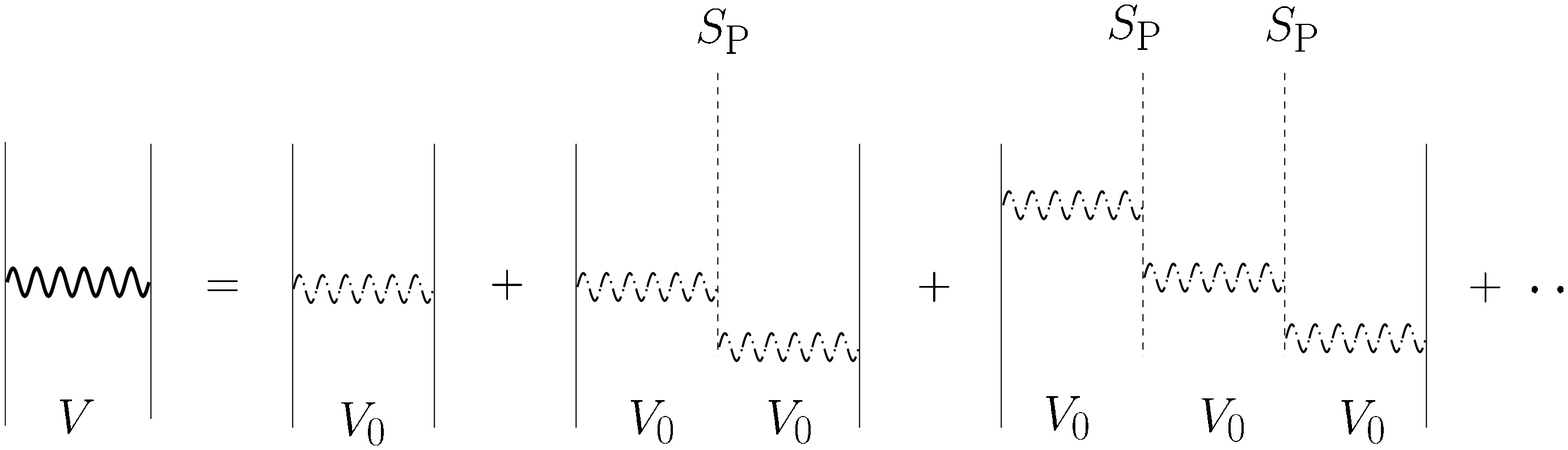}
\end{figure}

\begin{figure}
\epsfxsize=12cm
\epsffile{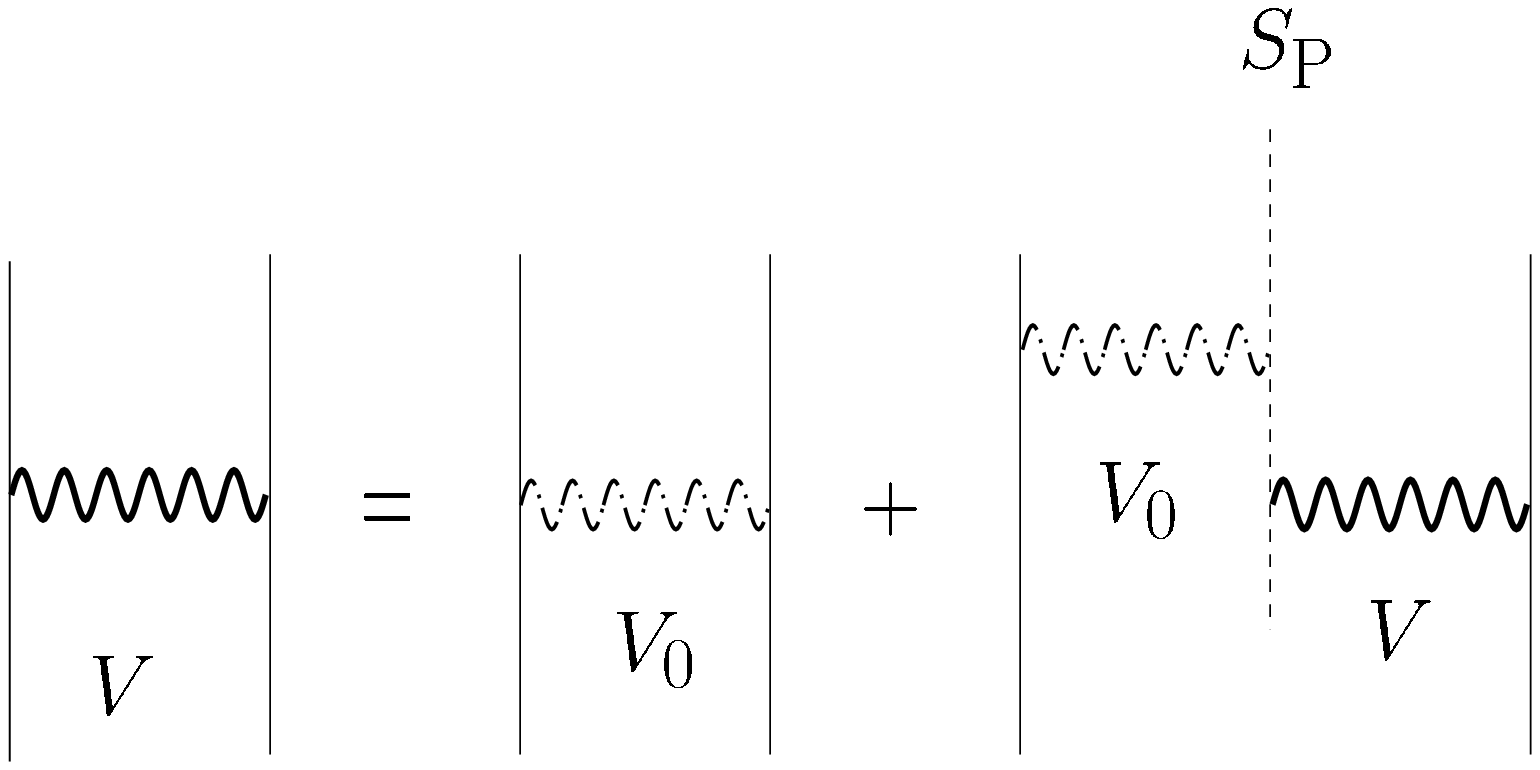}
\caption{b}
\end{figure}

\pagebreak
\begin{figure}
\epsfxsize=15cm
\epsffile{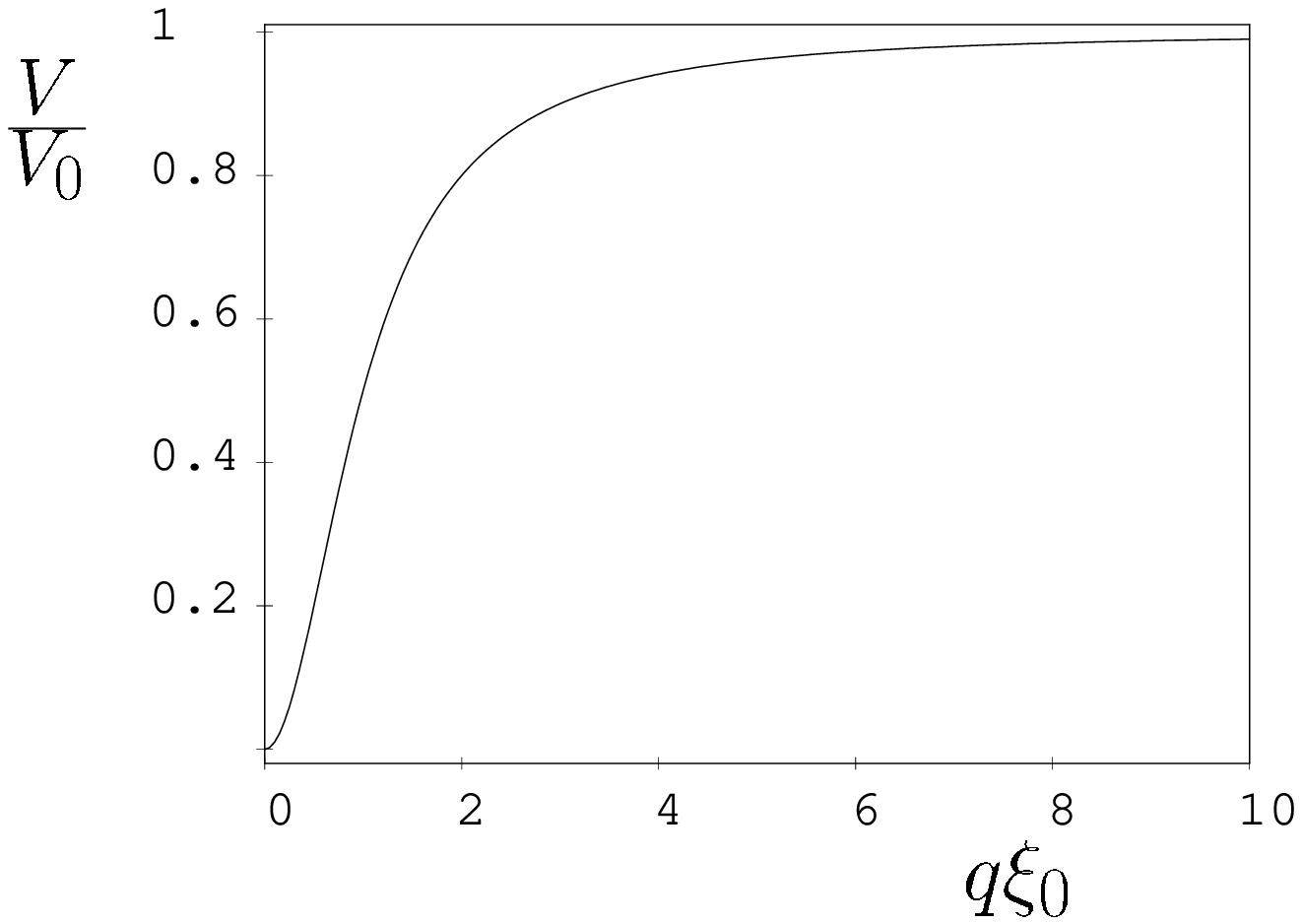}
\end{figure}

\end{document}